# Solar-induced thermal activity and stratification in pond water

James D. Brownridge

*Department of Physics, Applied Physics, and Astronomy, State University of New York at Binghamton, P.O. Box 6000 Binghamton, New York 13902-6000, USA*

**Abstract:** Ponds are universally used to store water for a large number of uses. With the increasing demand for more fresh water, ponds, lakes and reservoirs are likely to be constructed on a larger scale. We must understand the effects of environmental changes on fresh water if we are to most efficiently utilize this resource. This study undertakes to increase our understanding of the rate of thermal response of ponds and other bodies of water to every-day environmental changes. The central research agenda is to investigate how the temperature of pond water from top to bottom responds to the day/night cycle, changes in air temperature just above the surface, cloud conditions, and other sudden environmental changes. Data collection for this study spanned October 2007 to June 2011 and had a continuous time resolution of 50 seconds.

**Introduction:**

With a growing worldwide human population and a finite amount of fresh water, the increase occurrence of water shortages is a major concern. One way to make more of this finite resource available for human use is by building dams, reservoirs and ponds, which has been done for some time. Of course, such construction has and will continue to have major effects on the ecosystem, a problem that will not be addressed here. In this paper the high-resolution thermal activity of the pond, whereby the temperature was continuously measured every 50 seconds throughout the water column, is presented. Ruochuan Gu et al. (1996) used a resolution of 20 minutes in a similar study, whereas other studies, Abis, Karen L. *et al.*, (2006), Escobar, Jaime *et al.*, (2009) and Moss, Brian. (1969, used much longer times between data points. However, there appears to be no literature on long-term data collection with time intervals between data points of less than 20 minutes; the purpose of this year long study was to study thermal activity in pond water in response to rapid changes in the environment above the surface of the water. The data presented in this paper is presented in a graphic format to allow the reader to graphically analyze and interpret the data.

**Method:**

Temperature probes, PASCO Model PS-2135 thermistors sealed in glass test tubes with waterproof silicone, were positioned ~50 cm apart from top to bottom near the center of the pond. The test tubes were secured upside down to a weighted drop line to minimize water seepage into the tube, which would damage the thermistor. The top thermistor was approximately 10 cm below the surface of the water and held in place by a float and tie line across the pond, as shown in Fig. 1. The system was designed with the intent to collect data for as long as 3 years. The pond was oval in shape and ~107 meters long, ~38 meters wide and 3 meters deep at the site of the temperature probes. The top temperature probe was within ~10 cm of the surface of the water, and the bottom probe was in sediment at the bottom of the pond. The air temperature just above the surface of the water was recorded with a probe on top of the float. A solar cell was located at the edge of the pond, and its output was used to record sunrise, sunset, cloudy days and when a cloud passed between the sun and the pond on partly cloudy days. Data from each sensor were collected once per second for 50 seconds and then averaged and saved as one data point. Data was collected continuously using two "PASCO plorer GLX Data loggers." (http://www.pasco.com/) The pond is located near the city of Binghamton, New York, USA, at Latitude 42.030564 N and Longitude -75.882511 W. Water flows into the pond from a meadow and a spring that flows year round, as shown in Figs. 1 and 2. A typical example of how these sensors responded during the daylight is shown in Fig. 3. At the start of experiment during an equipment test run, it was noted that on clear days the air temperature began rising before the solar cell indicated the sunrise; therefore, the increase in air temperature was used as "time zero" when describing events in Fig. 3. The solar cell responded to sunrise approximately 91 minutes after the air responded. The temperature of the water began rising approximately 351 minutes after the air. Approximately 658 minutes after "time zero," a dark cloud moved between the sun and the pond. The cloud blocked direct sunlight from entering the water for approximately 73 minutes. Maximum blockage of direct sunlight occurred at (c), approximately 28 minutes after the cloud first started moving between the sun and the pond. The temperature of the air fell to the lowest point at (d), after 18 minutes, and the temperature of the water at all levels in the pond fell to the lowest temperature 12 minutes later. After the cloud passed and direct sunlight entered the water, the rise in temperatures followed a similar pattern. Daylight on August 3, 2008 officially lasted for 14.3 hr, and the solar cell recorded daylight as lasting 14.7 hours.

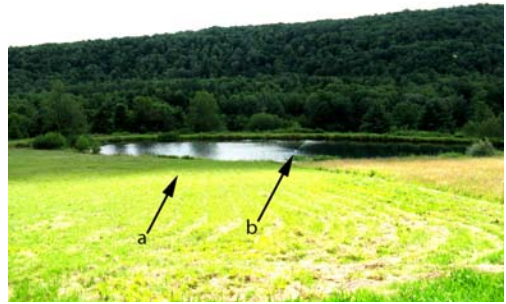

Fig 1. Photograph of the pond in late summer as a dense cloud passed between the sun and the pond. Arrows (a) and (b) point to the location of the passing shadow and the temperature sensors.

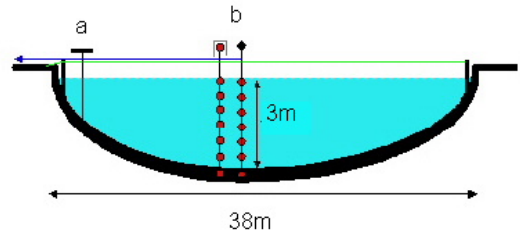

Fig. 2. Schematic diagram of the pond and sensors. (a) Solar cell used to monitor sunrise, sunset and cloud conditions. (b) Temperatures probes used to monitor water and air temperatures.

**Results and Discussion, Spring and Summer:**

First, thermal activity in the water over a period totaling more than a year will be looked at broadly.

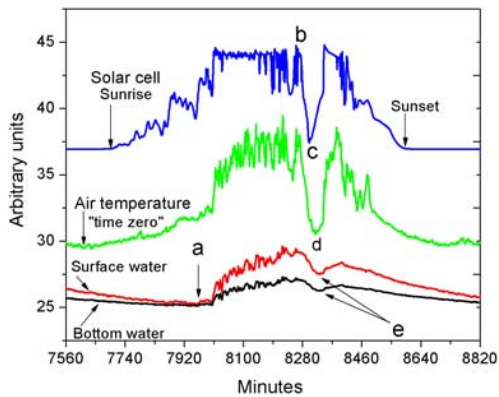

Fig. 3. Four of the sensors that best illustrate how the system responds to a major event. (a) Temperature of water starting to rise. (b) Cloud began blocking direct sunlight. (c) Maximum direct sunlight blockage. (d) Minimum air temperature during blockage. Other sensors are not shown here for clarity.

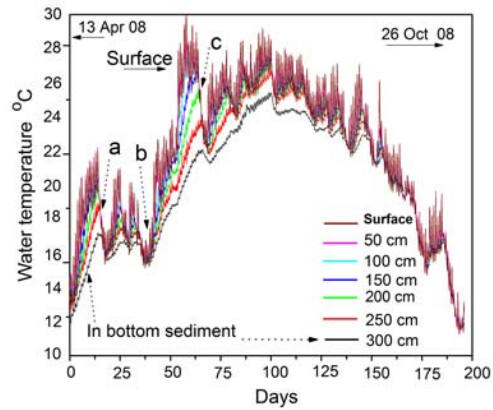

Fig. 4. Continuous temperature profile of the 3 m deep pond beginning in the spring on April the 13th 2008 and ending in the fall on October the 26th 2008.

In Fig. 4, thermal activity from mid-spring to early fall in the year 2008 is shown. In Fig. 5, thermal activity from fall to spring in the years of 2009 and 2010 is shown. As expected, many weather events occurred during this time, and several will be examined and discussed in detail.

Each spike at the top of the graph in Fig. 4 corresponds to one day. The height of the spike is the maximum temperature of the water at the surface of the pond on that day. The lower trace is the temperature of the sediment at the bottom of the pond. The graph traces between the lower trace and the top trace, indicating the temperature of the water at ~50 cm intervals from the bottom to the top of the pond.

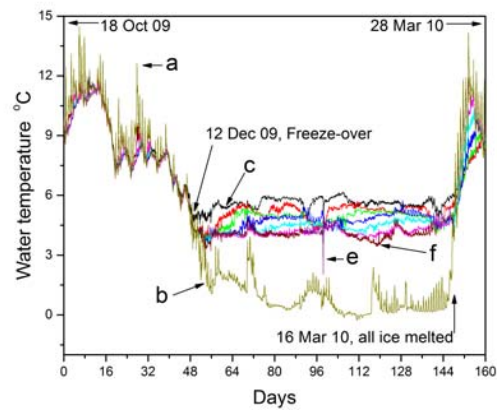

Fig. 5. Continuous temperature profile of the 3 m deep pond beginning in the fall on October the 18$^{th}$ 2009 and ending in the spring on March the 28$^{th}$ 2010. Note that on the 12$^{th}$ of December, the pond froze over and the bottom became warmer than the top and remained so for the rest of the winter.

April 23$^{rd}$ 2008 began as a partly sunny morning, with heavier clouds moving in by noon and dark clouds rolling in by early afternoon; this resulted in a sudden drop in air temperature just above the surface of the water and a corresponding drop in surface water temperature, as shown at (a) in Fig. 6. Notably, at (b), the temperature of the water 50 cm below the surface stopped increasing and remained nearly constant for the remainder of the day, up until after sunset, at (c). By contrast, the following day, April 24th, had a clear sky, and the temperature of the water at 50 cm continued to rise after sunrise until late afternoon. The weather on April 25th was similar to April 23rd. In the early spring during the late afternoon when the sun was low on the horizon and setting, the air temperature began to drop, and the temperature of the water at the surface also began falling and fell below that of the water ~1 meter deep, as shown at (c) in Fig. 6. The probe measuring the temperature at the bottom of the pond was omitted for clarity in Fig. 6. The trace marked (2) in Fig. 6 is the record of the temperature at 2.5 meters deep, and the trace marked (7) was at the surface.

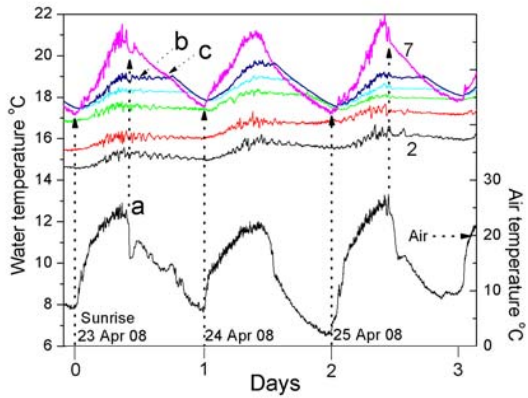

Fig. 6. Variation in water temperature at various depths in response to day/night cycles and changing weather conditions. From the top, (**7**) was the probe at the surface, next, probe (6) was 50 cm down, then (5) was 100 cm down, (4) was at 150 cm, (3) was at 200 cm, and (2) was 250 cm deep.

August 19th 2008 was a rainy day; August 20th was a bright, partly cloudy day; and the 21st, the last day shown in Fig. 7, was a very clear blue sky day.

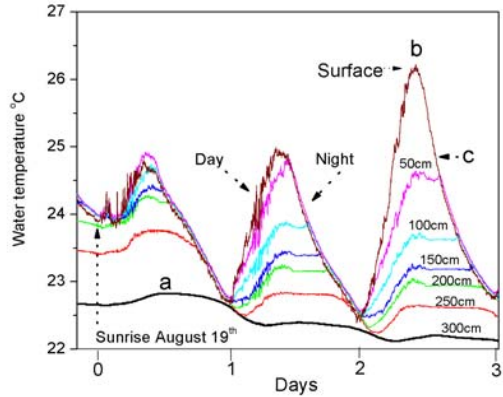

Fig. 7. Water temperature at various depths in response to day/night cycles and changing weather conditions.

There was no noticeable wind during this period of data collection. During this unique three-day period, the effect that clouds had on the thermal activity of the water at every depth in the pond could be seen. Fig. 7 (a) showed that the pond was slowly cooling; this was evident from the probe located in the sediment at the bottom of the pond. The pond had reached the maximum temperature in late July (see Fig. 4, around day 100). August 21st (b), was a crystal clear day with no wind. When this occurred, good temporary thermal stratification that quickly collapsed after sunset, as seen at (c) in Fig. 7, was noted. Long-term pond thermal stratification occurred in a step fashion beginning at sunrise each day and continued over many days, as shown in Fig. 8; however, it collapsed within 24 hours when direct sunlight was blocked by clouds during daylight hours. This affect was dramatic when the air temperature during the night fell significantly below that of the temperature of the water, see (a) and (c) in Fig. 4. and at (a) in Fig. 8 . During this study, there were two long periods in the early spring when thermal stratification extended from top to bottom for more than two or three days was observed. For most of the summer, thermal stratification at the bottom of the pond that lasted several months was observed.

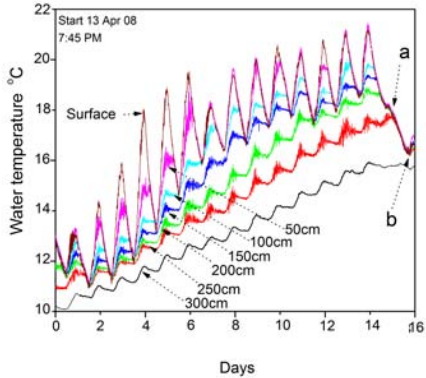

Fig. 8. Day-by-day step stratification of the pond in early spring and the collapse in one day at (a).

In the spring and summer, there was more jitter, short-term reversal in the direction of temperature changes ranging from several minutes to an hour, in the temperatures signals than there was at night. This was true on the calmest, sunniest day; see Figs. 7 and 9.

However, there was one period in May 2008 when the temperature jitter continued after sundown, as is shown in Fig. 10 on day 3. On partly cloudy days, clouds passing between the sun and the pond produced most of the jitter. On windy days, waves produced by the wind caused sunlight that on a calm day would have otherwise been absorbed by the water to be reflected away from pond; therefore, less solar energy went into the water. On days when the sky was clear and the wind was calm, a nearly smooth temperature increase throughout the water column during daylight and a very smooth fall in temperature at night was observed; see (b) in Fig. 7.

A question arose: how much of the jitter in temperature was caused by wind and how much was caused by interrupted solar energy input from clouds. Fig. 9 shows two days in June 2008, the 9th and the 10th. Both days were mostly clear in the morning and mostly cloudy in the afternoon, with intermittent clouds passing between the sun and the pond and a light wind gust. On June 10th at ~6:25 PM, a fast-moving thunderstorm passed through with 50 mph wind gusts. By 7:45 PM, the storm had passed and the sun was out. The higher flutter seen in trace (2) during and after the storm was attributed to the storm. The effect was much stronger near the bottom of the pond than nearer to the surface, which may be because of wave motion that caused the float to move up and down, thereby causing the probe to move into warmer and cooler water near the bottom of the pond. The trace labeled (0) is air temperature and the trace labeled (1) is the temperature probe in the sediment.

The turbidity rose to ~5.2 NTU in early June, 2008 and fell to near 0.1 NTU for the rest of the summer.

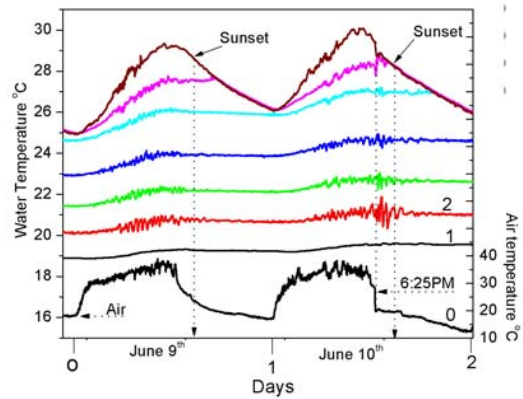

Fig. 9. Effect of a fast-moving storm. On June 10, 2008, a fast-moving thunderstorm passed over the pond at 6:25 PM, with rain and wind gust up to 50 mph.

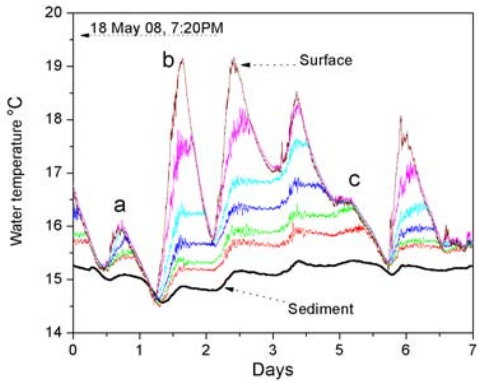

Fig. 10. Mostly cloudy day (a), clear sky sunny day (b) and (c) a rainy day all day.

## Results and Discussion, Fall and Winter:

Once the pond began cooling in late summer, thermal stratification became a single day event that occurred on bright sunny days, as shown in Fig 11. Note that for most of this period, the night-time temperature of the water at the surface of the pond cooled to below the temperature at the bottom, regardless of how high the daytime temperature rose. During an overcast rainy day, the pond was at nearly the same temperature at all depths (a) and it continued cooling day and night, as shown in Fig. 11 at (a) and (b).

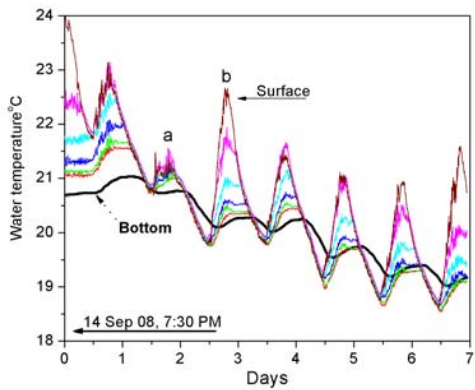

Fig. 11. Day/night cycles when the warmest water is at the bottom of the pond at night.

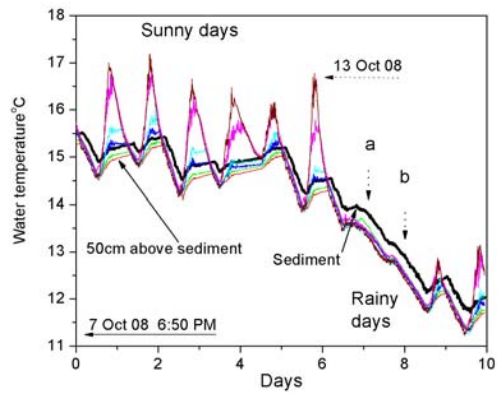

Fig. 12. Thermal activities in the water prior to pond freeze-over.

During the fall prior to freeze-over, thermal stratification developed in the water near the bottom of the pond and remained stratified for several days. At the same time, the water in the upper part of the pond went through day/night periods of stratifying during the day and collapse at night, as shown on the sunny days in Fig. 12. On rainy days, the water above approximately 50 cm above the bottom of the pond was around the same temperature and showed a cooling trend, see (a) and (b) days 7 and 8 in Fig. 12. Additionally, the coldest water during days 6-8 was all the water 50 cm above the bottom of the pond.

The process of the pond freezing over often took several days; however, when the freeze finally occurred, the pond became thermally active near the surface, as shown in Fig. 13. The effect of the day/night cycle deep in the water was apparent in the water just under the ice after freeze-over, as shown in Fig. 14(A). Fig. 14(B) is a graphic display of the air temperature and the solar cell response to dally cloud conditions.

In Fig. 14, day 1 was mildly overcast, day 2 was heavily overcast, and days 3

and 4 had scattered clouds. Day 13 in Fig. 14(B) was a clear day. The solar cell output was a reflection of the type of weather a on a particular day. The generally decreasing temperature of the water near the surface was an indication that the ice sheet was thickening and had almost reached the depth of the top temperature probe by day 14. The temperature of the water under the ice sheet rose with each sunrise on partly cloudy or sunny days. On cloudy days, a temperature increase with sunrise was not as clear. During the spring and summer and before freeze-over, a consistent rise in air temperature before sunrise on partly cloudy or sunny days was observed. The effect of sunrise on the air temperature during winter was not as consistent or dramatic as it was during summer; see Fig. 14(B). The only time long-term stable pond thermal stratification was observed was in mid-winter when the air temperature remained near or below 0°C for an extended time, as shown in Fig. 15.

On the night of January 20, 2010, a warm front passed through producing ~5 cm of rain in less than 24 hours. The pond was covered with a sheet of ice > than 8 cm.

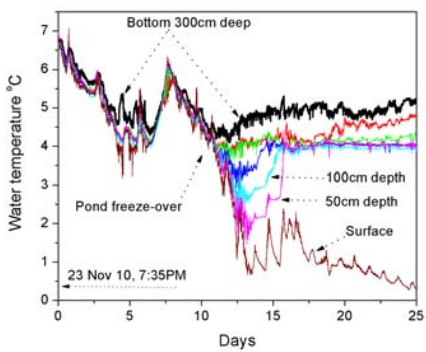

Fig. 13. Pond freeze-over followed by winter thermal stratification.

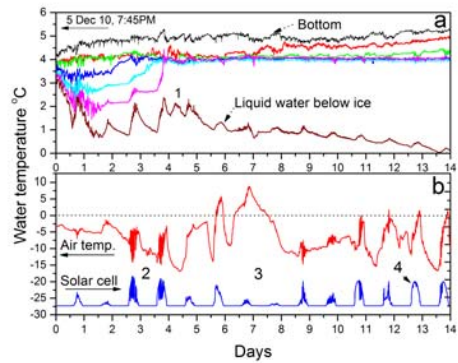

Fig. 14. Days after freeze-over when the pond was covered with a thin sheet of ice. (A) Sensors in the water. (B) Sensor in the air.

Cold runoff from the frozen meadow above the pond (shown in Fig. 1) flowed to the pond and then over and under the ice sheet. The temperature of the water flowing into the pond was close to 0°C because the meadow was covered with snow before it starting raining. Because this incoming water was colder and heavier than the water currently at the bottom of the pond, it sank to the bottom, thereby temporally lowering the temperature of the water at the bottom, as shown in Fig. 16. This unique rain event produced the only major thermal event in the water during freeze-over.

Water just under the ice sheet showed a smooth rise in temperature following the sunrise for every day, except the day of the rainstorm. As the ice sheet began melting in late winter, the rise in temperature of the water just under the ice was smooth and dramatic on sunny days, as shown by (a) in Fig. 17. The temperature of the water at the bottom began to show a response to the day/night cycle once the ice thinned and there was no longer snow on the ice. When the pond was free of ice on the morning of March 16, 2010, the daily thermal stratification resumed, and the pond turned over thermally.

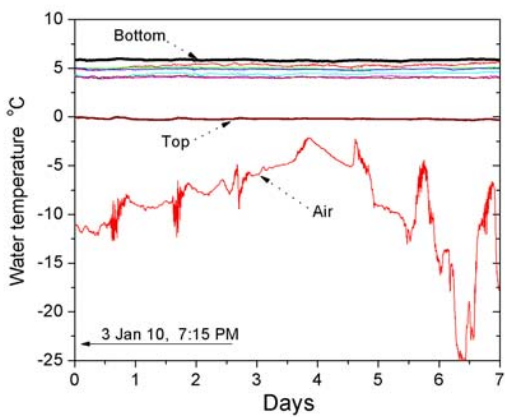

Fig. 15. Stable pond stratification under a sheet of ice covered with snow.

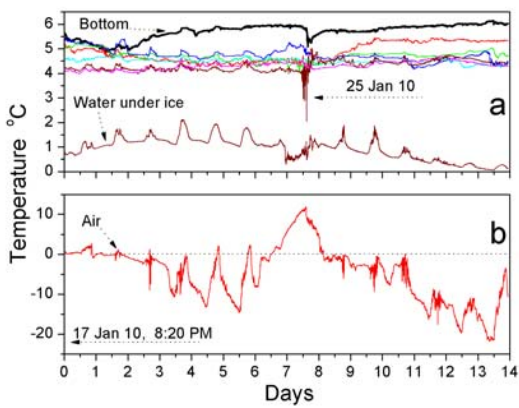

Fig. 16. Effect of a 20°C temperature rise beginning on the night of January 23$^{rd}$ and a 5 cm rainfall January 25, 2010.

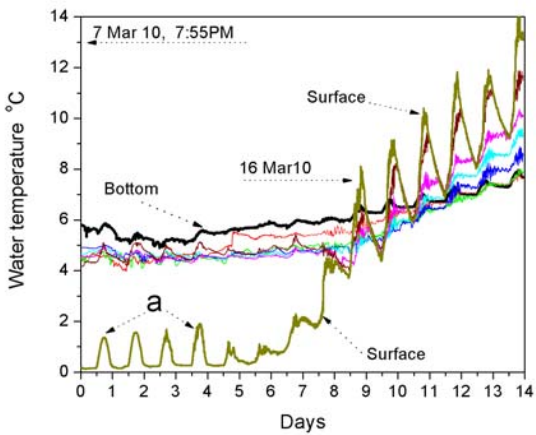

Fig. 17. Transitioning from winter to spring.

## Conclusion:

Direct sunlight on a small body of water was a major factor affecting the daily and intraday changes in the temperature of the water throughout the water column. A succession of bright, sunny days produced thermal stratification in a step-wise fashion, building from the top down, as long as there were uninterrupted sunny days. One cloudy day may cause thermal stratification to collapse, and two successive cloudy days will cause a complete collapse of thermal stratification. Turbulence at the surface or waves produced by wind reduced the amount of solar energy entering the pond by continuously changing the reflective properties of the surface. The net effect was less energy input on windy days, more jitter, a smaller temperature increase and more thermal activity in the water column. Temperature jitter was always present, i.e. thermal activity during the daylight affected temperature traces because of clouds and surface turbulence, whereas during the nighttime traces were generally smooth on both windy and calm nights.

## Acknowledgement:

I thank S. M. Shafroth and M. D. Stephens for many helpful discussions and suggestions.

jdbjdb@binghamton.edu